\documentclass[english]{aastex}
\usepackage[T1]{fontenc}
\usepackage{graphicx}

\makeatletter

\providecommand{\tabularnewline}{\\}

\slugcomment{}
\shorttitle{}
\shortauthors{}

\makeatother

\usepackage{babel}

\begin{document}

\title{\emph{MOST} Space Telescope Photometry of the 2010 January Transit
of Extrasolar Planet HD80606b}

\author{Jessica E. Roberts}

\affil{San Jose State University}

\affil{Department of Physics and Astronomy, Room 148}

\affil{San Jose, CA 95192-0106}

\email{jessica.roberts@sjsu.edu}

\author{Jason W. Barnes}

\affil{University of Idaho}

\author{Jason F. Rowe}

\affil{NASA Ames Research Center}

\author{Jonathan J. Fortney}

\affil{University of California - Santa Cruz}
\begin{abstract}
We present observations of the full January 2010 transit of HD80606b
from the Canadian microsatellite, Microvariability and Oscillations
of Stars (MOST). By employing a space-based telescope, we monitor
the entire transit thus limiting systematic errors that result from
ground observations. We determine measurements for the planetary radius
($R_{p}=0.987\pm0.061$$R_{Jup}$) and inclination ($i=89.283^{o}\pm0.024$)
by constraining our fits with the observed parameters of different
groups. Our measured mid-transit time of 2455210.6449$\pm$0.0034
HJD is consistant with the 2010 Spitzer results and is 20 minutes
earlier than predicted by groups who observed the June 2009 transit.
\end{abstract}


\section{Introduction}

With its unusually high eccentricity of $e=0.93$ and long orbital
period of 111 days, HD80606b challenges our understanding of exoplanets.
This planet was discovered by \citet{2001A&A...375L..27N} via radial
velocity techniques in which they determined a minimum planetary mass
of 3.9 M$_{Jup}$. The occultation of HD80606b was observed with the
Spitzer Space Telescope eight years later indicating that the inclination
of the planet is greater than 89$^{o}$\citep{2009Natur.457..562L}.
The existence of an occultation indicated that there was a 15\% chance
that this planet would also have a visible primary transit. 

Despite the low probability, several independent groups detected the
February 2009 primary transit from ground-based observatories \citep{2009MNRAS.396L..16F,2009ApJ...698..558G,2010MNRAS.406.1146H,2009arXiv0906.5605P,2009A&A...498L...5M}.
The full transit was later observed by \citet{2009arXiv0907.5205W}
who used multiple ground-based observatories to sew together complete
coverage. 

\citet{2009arXiv0907.5205W} also confirmed a spin-orbit misalignment
of 53$^{o}$ that was first predicted by \citet{2009A&A...498L...5M}.
It is possible that the high eccentricity and tilted orbit of HD80606b
are the result of the Kozai mechanism induced by the system's binary
stellar companion HD80607 \citep{2003ApJ...589..605W}.

Since the 12 hour duration of HD80606b's transit presents a difficulty
for accurate ground-based observations, \citet{2010A&A...516A..95H}
employed the combination of Spitzer photometry and $SOPHIE$ spectroscopy
to continuously observe the full January 2010 transit. \citet{2010A&A...516A..95H}
discovered their mid-transit time to be approximately 24 minutes earlier
than the value predicted by \citet{2009arXiv0907.5205W}. In addition,
\citet{2010A&A...516A..95H} found their $R_{p}/R_{*}$ value differed
by 3$\sigma$ compared to the results obtained by both \citet{2009arXiv0906.5605P}
and \citet{2009arXiv0907.5205W}. 

\citet{2010ApJ...722..880S} launched a multiple-site ground-based
observation of the same transit in an attempt to confirm the \citet{2010A&A...516A..95H}
timing discrepancy. Their midtransit time was 12 minutes earlier than
predicted and within 1.3$\sigma$ of \citet{2010A&A...516A..95H}.

We observe the January 2010 transit of HD80606b with the Microvariability
and Oscillations of Stars (MOST) earth-orbiting optical telescope
with the goals of: (1) determining a mid-transit time using both \citet{2009arXiv0907.5205W}
and \citet{2010A&A...516A..95H} orbital parameters and (2) derive
an independent set of results for the planetary radius, inclination,
and mid-transit time. Our observations by the MOST telescope are described
in $\S$2, our light curve modeling is detailed in $\S$3, and we
present and discuss our results to both goals mentioned above in $\S$4.
We conclude this paper in $\S$5.

\section{Observations}

\label{sec:observations}

The Microvariability and Oscillations of Stars (MOST) is a microsatellite
with a 15 cm aperture photometer designed for asteroseismology, but
is well-suited to observe short period exoplanets around solar type
stars \citep{2006ApJ...646.1241R}. The satellite employs a broadband
visible filter centered at 525 nm. MOST is in a low polar, sun-synchronous
orbit that permits for a continuous viewing zone (CVZ) of stars between
the declination of -19$^{o}$ and +36$^{o}$. HD80606b, however, has
a declination of +50$^{o}$ placing it outside of the CVZ. Therefore,
we experience short blackouts in our data as Earth passes in between
MOST and HD80606. We are unable to observe the January 8th, 2010 secondary
eclipse of HD80606b due to an inconvenient instrumentation issue that
led to a 12 hour gap in our data.

We observe HD80606b starting from 24548677.97302083 HJD and continuing
to 2454877.85480324 HJD (January 7th to January 19th, 2010) with an
exposure time of 80.40 second integrations. Since HD80606 is a magnitude
9.06 star, we use the direct imaging mode of the CCD and employ simple
aperture photometry with a field of view of 20 x 20 pixels. We capture
images of both HD80606 and HD80607, but we are able to constrain HD80606
within a 3 pixel radius compared to the binary stars' separation of
10 pixels.

Stray light for MOST varies from 100 ADU to 1500 ADU as it passes
over the polar caps and day/night boundary of Earth. We remove data
that has a background level above 800 ADU (about 100,000 points).
At this lower level of stray light, our photometric error is equal
to the expected Poisson noise limit. For a detailed procedure of standard
observation and data reduction of MOST data, see \citet{2006ApJ...646.1241R}. 

\begin{figure*}
\includegraphics[width=0.75\textwidth]{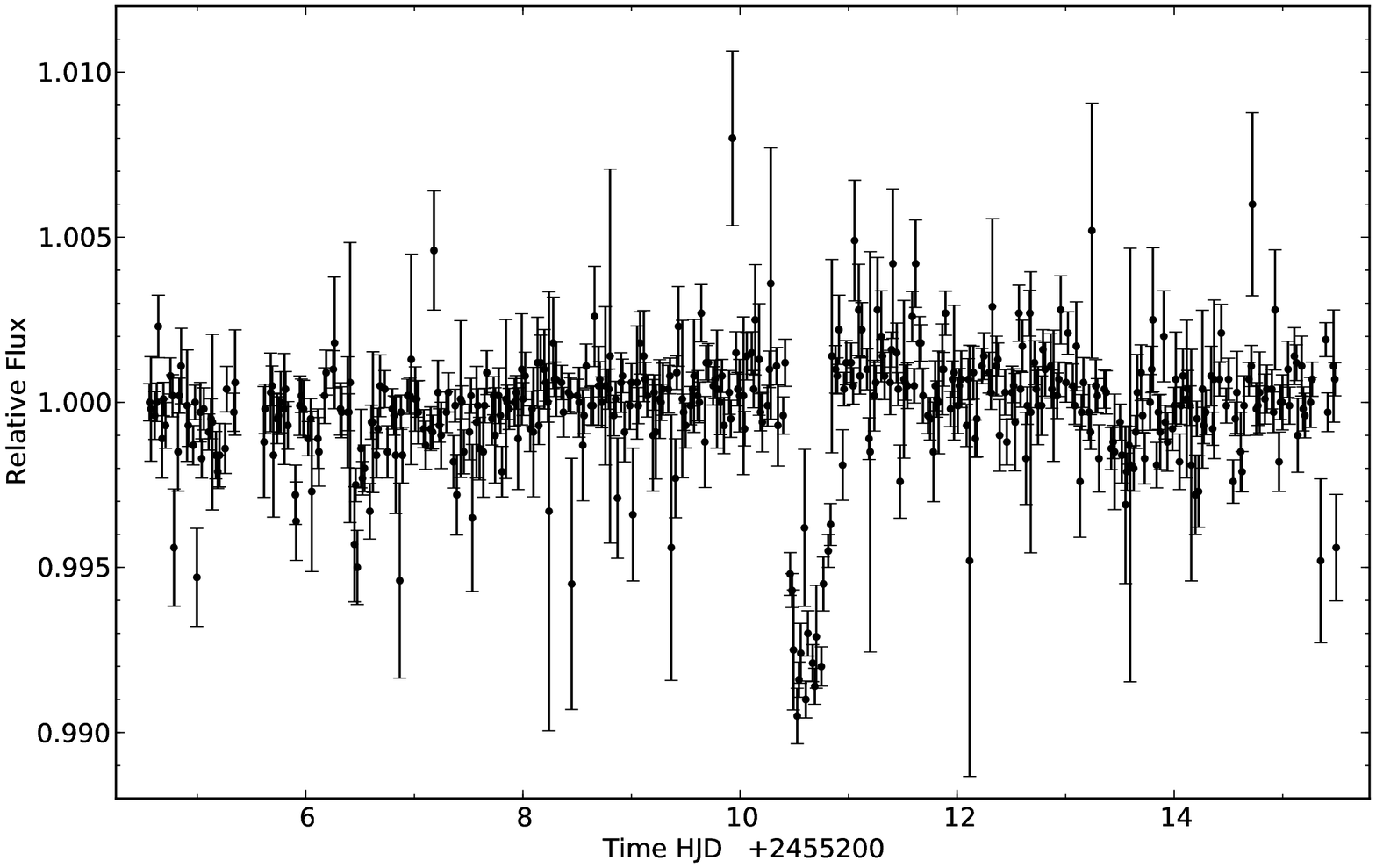} \caption{Here we show the MOST photometry full data set, including errors.
The photometric data are binned to one point every 15 minutes . \label{figure:fulldata}}
\end{figure*}

\begin{figure*}
\includegraphics[width=0.75\textwidth]{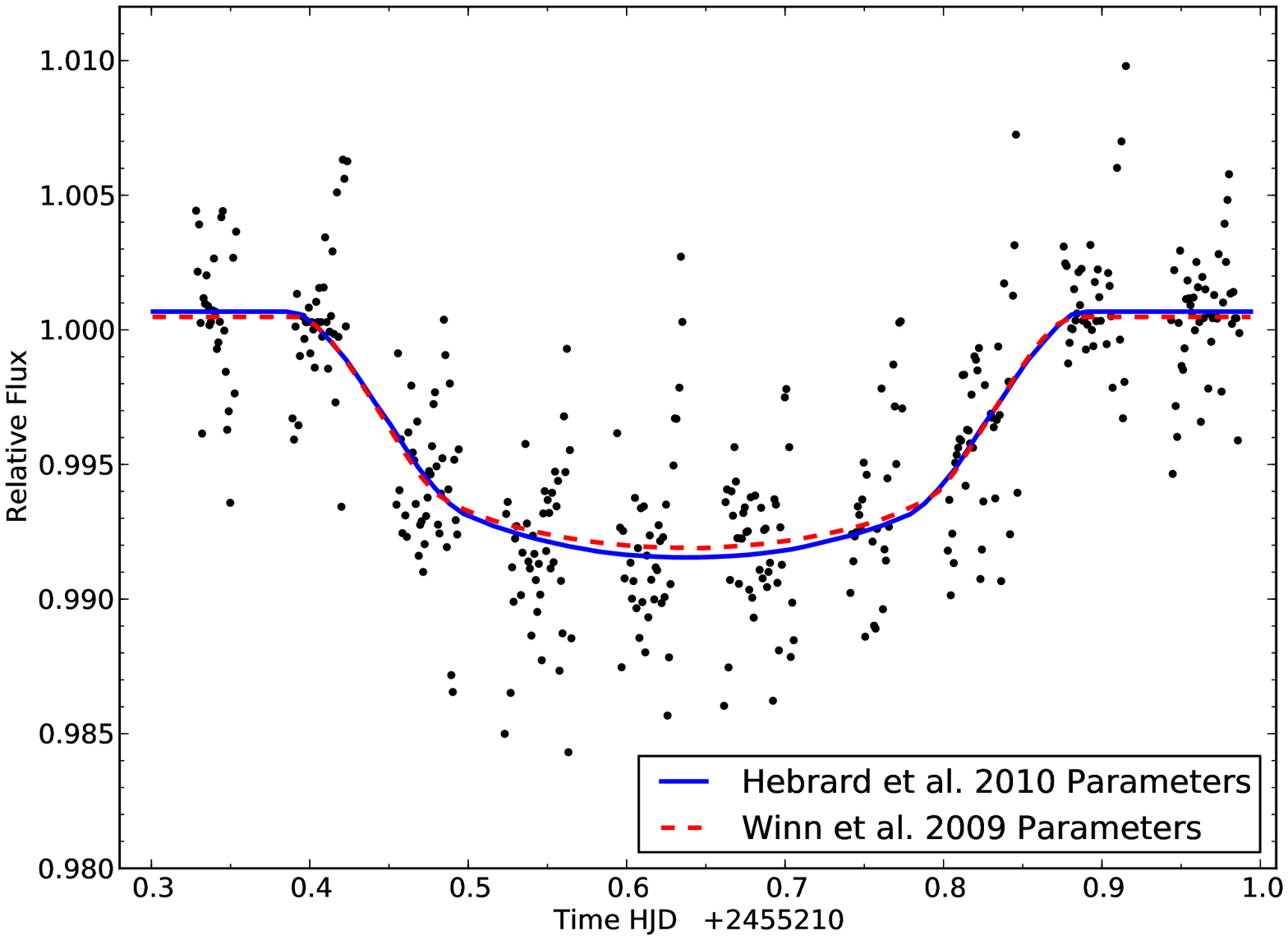} \caption{Unbinned close-up of the MOST transit data. The two lines represent
our two best fits. The blue line is our best fits constrained by Hebrard
et al. 2010 parameters, whereas the red dashed line is the best fit
constrained by Winn et al. 2009 parameters. \label{figure:fits}}
\end{figure*}

Figure \ref{figure:fulldata} shows our full photometric time series,
with points binned every 15 minutes for clarity. By binning the data
the gaps are not as apparent; however, we fit the full unbinned transit
(Section 3). Due to the binning effect, the transit in Figure \ref{figure:fulldata}
appears asymmetric. A close up of the unbinned transit in Figure \ref{figure:fits}
demonstrates that within the error of our measurement, the transit
is symmetric and we are not forced to make necessary modifications
for an asymmetric effect. The large gap occurring at 24552005.5 is
due to the instrumentational error discussed above.

\section{Lightcurve Modeling}

We use the fitting process described in \citet{oblateness.2003}.
With high eccentricity, the transit duration is correlated to the
inclination and argument of periapsis of the planet's orbit. The transit
ingress and egress times also experience a several minute asymmetry,
with the ingress duration shorter than the egress in the case of HD80606b.
We account for these variations by explicitly incorporating orbital
eccentricity in to our code, as detailed in \citet{2007PASP..119..986B}. 

In this routine, we do not account for the impact parameter $b$ as
this parameter is directly related to the inclination in an eccentric
orbit by\begin{equation}
b=\left(\frac{a}{R_{*}}\right)\left(\frac{1-e^{2}}{1+e*sin(\omega)}\right)cos(i)\label{equation:impactparameter}\end{equation}
 By fitting for the inclination alone, we accelerate the fitting algorithm.

We fit the out-of-transit flux dynamically in our model instead of
assuming a value. Out-of-transit points that fall more than 3$\sigma$
outside of this photometric baseline are removed. Our best fit transit
model is determined by minimizing$\chi^{2}$ through a Levenberg-Marquardt
fitting algorithm. 

As our data is not precise enough to fit for the limb darkening parameters
combined with our other fitted parameters, we hold both coefficients
constant. Our limb darkening parameters are established by using the
defined MOST quadratic parameters of HD80606 with a $T_{eff}=5645K$
and $log$ $g=4.50$ \citep{2001A&A...375L..27N}. These coefficients
are manipulated to fit our model by using the limb darkening method
highlighted in \citet{2001ApJ...552..699B}. Our first coefficient,
designated as $c_{1}=0.742$, represents the magnitude of darkening
while our second coefficient, $c_{2}=0.458$, accounts for the curvature
of the transit \citep{oblateness.2003}. 

We do not account for any other systematic effects or correlated noise.
Therefore, we assume all errors to be random with a Gaussian distribution.
See \citet{oblateness.2003,2007PASP..119..986B} for additional details
regarding our model. 

We execute two different model fits: the first model is used to verify
the mid-transit time reported by \citet{2010A&A...516A..95H} and
a second to determine our own independent parameters. In our effort
to reproduce the \citet{2010A&A...516A..95H} mid-transit time, we
hold all their parameters constant and fit solely for the baseline
flux and mid-transit time. We repeat this procedure using \citet{2009arXiv0907.5205W}
parameters for comparison to their predicted mid-transit time. For
clarity, we will denote our timing results with either a superscript
TH for \citet{2010A&A...516A..95H} and TW for \citet{2009arXiv0907.5205W}
when discussing out results in Section \ref{sub:transittime}.

In establishing our independent best-fit parameters, we determine
that our data are only adequate to fit for four parameters: planetary
radius $(R_{p})$, mid-transit time $(T_{mid})$, inclination $(i)$,
and the baseline flux. We constrain our two fits with the parameters
from \citet{2010A&A...516A..95H} and \citet{2009arXiv0907.5205W}
for stellar radius, stellar mass, period, eccentricity, argument of
periapsis, and semi-major axis. We analyze these results in Section
\ref{sub:Fit-Parameters}. 

\section{Discussion}

\subsection{Mid-Transit Time Fit Only \label{sub:transittime}}

In this section, we discuss our mid-transit time results of HD80606b
with a goal of verifying the early mid-transit time discovered by
\citet{2010A&A...516A..95H}. In order to measure a mid-transit time
appropriate for comparison, we hold \citet{2010A&A...516A..95H} orbital,
stellar, and planetary parameters constant and fit only for the mid-transit
time and baseline flux.

Our resulting mid-transit time (denoted with the superscript TH) is
20 minutes earlier than predicted by \citet{2009arXiv0907.5205W}.
For comparison, \citet{2010A&A...516A..95H} found a mid-transit time
24 minutes (17$\sigma$ difference) earlier than the value predicted
by \citet{2009arXiv0907.5205W}.

Orbital parameters have a direct effect on mid-transit times, especially
in a highly eccentric orbit. Since \citet{2010A&A...516A..95H} determined
a separate set of orbital parameters compared to \citet{2009arXiv0907.5205W},
we also fit the MOST lightcurve using the latter's parameters for
the mid-transit time. Even with the different orbital parameters,
we determine a mid-transit time (denoted with the superscript TW)
20 minutes early as well. 

\begin{figure*}
\includegraphics[width=0.75\textwidth]{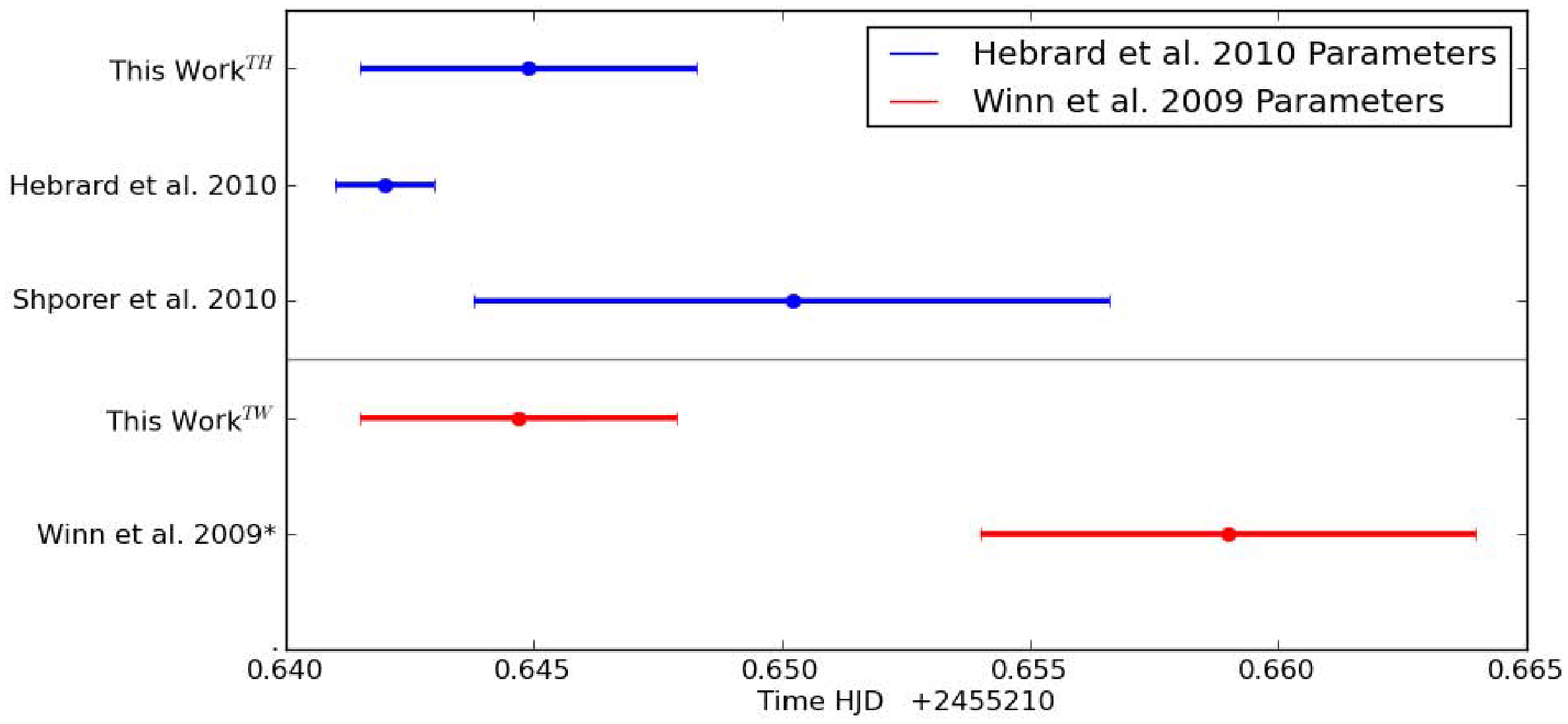} \caption{Five different mid-transit timing results from different groups with
errors included. The top three results in blue were calculated using
Hebrard et al. 2010 parameters. The bottom two results in red were
calculated using Winn et al. 2009 parameters. \label{figure:timing}}
\end{figure*}

We conclude that the difference in mid-transit times between the two
groups is not dependent on their varying orbital parameters. Furthermore,
we determine that our results have less than a 1$\sigma$ discrepancy
from the mid-transit time determined by \citet{2010A&A...516A..95H}.
A visual comparison of each mid-transit time with uncertainties is
displayed in Figure \ref{figure:timing}. We also highlight the mid-transit
timing results from an independent ground-based group, \citet{2010ApJ...722..880S},
in Figure \ref{figure:timing} who observed the 2010 January transit.
Our overall timing results are also listed in Table \ref{table:times}.

Our mid-transit time agrees with the \citet{2010A&A...516A..95H}
result to a less than 1$\sigma$ difference. We also find that our
mid transit time has 1.6$\sigma$ difference with the result determined
by \citet{2010ApJ...722..880S} and a greater than 3$\sigma$ difference
from the predicted transit time of \citet{2009arXiv0907.5205W}. In
fact, the only result that agrees with the \citet{2009arXiv0907.5205W}
predicted time (within 1$\sigma$) is \citet{2010ApJ...722..880S}.
They note, however, that due to the difficulty of ground based observations,
the uncertainty in their measurement is large. \citet{2010ApJ...722..880S}
still find a mid-transit time of 12 minutes earlier than \citet{2009arXiv0907.5205W}
predicted, a large discrepancy regardless of the uncertainty. 

One possible explanation for the large difference between the mid-transit
times of this work and \citet{2009arXiv0907.5205W} is the possibility
of a perturbing body in the system creating a transit timing variation
(TTV). However, \citet{2001A&A...375L..27N} show that if there is
another body perturbing this system, it is not detectable via radial
velocity measurements. \citet{2003ApJ...589..605W} also argue that
while another planet may have formed around HD80606, no Jupiter-mass
planet could lie in a stable orbit between 0.05 to 100 AU. They further
show that no Earth-mass planet could exist between 1 and 20 AU. As
TTVs are sensitive even to small mass objects, there is the possibility
of the existence of a moon around HD80606b. Contrary to this, \citet{ExtrasolarMoons}
show that no moon massive enough to create a transit timing variation
could exist in a stable orbit around the highly eccentric planet.

With the combination of 5 different mid-transit timing results, the
possible mid-transit time of HD80606b spans over 24 minutes. The large
discrepancy between each group's result demonstrates the mid-transit
timing issue is unlikely due to an astrophysical event. Furthermore,
it is of interest that both this work and \citet{2010A&A...516A..95H}
(both space-based observations) agree within less than 1$\sigma$
difference of an early mid-transit time, while the ground-based results
determine a mid-transit time several minutes later. We therefore propose
the hypothesis that the difference in mid-transit time derives from
systematic errors that depend on the method used to observe HD80606b. 

\begin{table*}[htbp]
\begin{tabular}{|l|c|c|}
\hline 
 & Mid-Transit Time  & $\chi_{\mathrm{r}}^{2}$\tabularnewline
\hline 
MOST (This Work) $^{TH}$  & $2455210.6449\pm0.0034HJD$  & 1.11735\tabularnewline
\hline 
Spitzer (Hebrard et al. 2010)  & $2455210.6420\pm0.0010HJD$  & \tabularnewline
\hline 
Ground-Based (Shporer et al. 2010)  & $2455210.6502\pm0.0064HJD$  & \tabularnewline
\hline
\hline 
MOST (This Work) $^{TW}$  & $2455210.6447\pm0.0032HJD$  & 1.11436\tabularnewline
\hline 
Ground-Based (Winn et al. 2009{*})  & $2455210.6590\pm0.0050HJD$  & \tabularnewline
\hline
\end{tabular}

{*}Predicted

\caption{Our mid-transit times for the January 2010 transit. {}``MOST$^{TH}$''
designates mid-transit time using parameters determined by Hebrard
et al. 2010. {}``MOST$^{TW}$'' represents the mid-transit time
determined with Winn et al. 2009 parameters. $\chi{}_{r}{}^{2}$ shows
the reduced $\chi^{2}$of our best fit. \label{table:times}}
\end{table*}

\subsection{Complete Transit Fit Parameters\label{sub:Fit-Parameters}}

We determine two independent sets of parameters for HD80606b, which
we will discuss and compare in this section. Figure \ref{figure:fits}
shows MOST's HD80606b transit lightcurve with the data unbinned and
each point representing 80.40 seconds of observation. The large gaps
between each cluster of data is the result of Earth passing between
MOST and HD80606 \citep{2003PASP..115.1023W}. Unfortunately, two
of these blackouts occur during large parts of the transit ingress
and egress. In order to minimize the effect of these blackouts on
our fit, we hold our limb darkening coefficients constant during the
entire fitting routine. 

We further minimize the effects of scatter and blackouts in our data
by electing to fit for only four parameters. These four parameters
are: planetary radius ($R_{p}$), inclination ($i$), mid-transit
time ($T_{mid}$), and the out-of-transit baseline flux. In this section,
we will denote our results with either a superscript FH or FW, for
\citet{2010A&A...516A..95H} or \citet{2009arXiv0907.5205W} respectively,
based on which of the groups parameters were used (e.g.: $R_{p}^{FH}$
or $R_{p}^{FW}$; $i^{FH}$ or $i^{FW}$, etc.).

We perform two separate fits using parameters obtained by \citet{2010A&A...516A..95H}
and \citet{2009arXiv0907.5205W} as fixed constraints. These parameters,
which we hold constant while fitting, are listed in Table \ref{table:constantparameters}.
Our two sets of results (corresponding to the two different sets of
fixed parameters) for the planetary radius, inclination, and mid-transit
time are listed in Table \ref{table:fitparameters} along with the
results of both \citet{2010A&A...516A..95H} and \citet{2009arXiv0907.5205W}
for comparison. We also report the $\chi_{r}^{2}$ of each of our
fits. However, the difference between the two $\chi_{r}^{2}$ is minimal.
This leads us to conclude that while our results determined by using
the \citet{2010A&A...516A..95H} constant parameters yields a smaller
$\chi_{r}^{2}$, the accuracy of our two fits is statistically comparable.
This agreement is demonstrated in Figure \ref{figure:fits} as we
plot our two best fits against each other. The solid blue line represents
our best fit employing \citet{2010A&A...516A..95H} constraints while
the red dashed line is our best fit using \citet{2009arXiv0907.5205W}
parameters. 

We assume that the $R_{p}/R_{*}$ ratio should be the same for both
our fits as the transit depth of our data does not change. We find
that our $(R_{p}/R_{*})^{FW}$ and $(R_{p}/R_{*})^{FH}$ measurements
are separated by 1.2$\sigma$. Regardless of the difference in our
stellar radius constraint, we still achieve statistically similar
results. Our $R_{p}/R_{*}$results straddle \citet{2010A&A...516A..95H}
result. We also find that our $(R_{p}/R_{*})^{FW}$ is the smallest
ratio of the set of results whereas \citet{2009arXiv0907.5205W} determined
the largest $R_{p}/R_{*}$ for HD80606b with a difference of 3$\sigma$.

The orbital inclination is directly dependent on the stellar radius,
making it difficult to compare our two $i^{FH}$ and $i^{FW}$ results.
We find it interesting that both our $i^{FH}$ and $i^{FW}$ results
are exactly 1$\sigma$ larger than the inclinations determined by
\citet{2010A&A...516A..95H} and \citet{2009arXiv0907.5205W} respectively.
This is likely due from the blackouts occurring in the ingress and
egress of our data, which cause a difficulty in fitting the exact
curvature of the transit. We do not report an independent estimation
of the impact parameter, as we do not fit for this variable. Instead,
we calculate $b$ using Equation \ref{equation:impactparameter} to
allow for comparison and find that our $b^{FH}$ is slightly larger
than our $b^{FW}$ but less than 1$\sigma$ difference. 

Our mid-transit times in both of these complete fits differ from the
timing only fits listed in Table \ref{table:times} because they are
the best fit times based on the other fitted parameters. As we fit
for multiple variables, the resulting errors on these times are higher
than the ones listed above. We still find a mid-transit time earlier
than predicted; however, the spread between our $T_{mid}^{FW}$ and
$T_{mid}^{FH}$ results is a 3 minute difference, even though they
have a less than 1$\sigma$ discrepancy. Interestingly, our $T_{mid}^{FW}$
is the earliest of all our mid-transit times, including those reported
in Section \ref{sub:transittime}, leading us to believe the mid-transit
time dependence on the other \citet{2009arXiv0907.5205W} parameters
is not causing the transit timing variation. Both of our $T_{mid}^{FW}$
and $T_{mid}^{FH}$ results are within 1$\sigma$ of \citet{2010A&A...516A..95H}
mid transit time and are at least 1.2$\sigma$ away from \citet{2009arXiv0907.5205W}. 

We verify our mid-transit time by fitting public spectroscopic data
from \citet{2010A&A...516A..95H} and our photometric data simultaneously.
While the quality of our transit leads to large uncertainties in most
of our parameters, we determine a mid-transit with less than 1$\sigma$
difference to our $T_{mid}^{FW}$ . We further confirm the period,
eccentricity and longitude of periapsis determined by \citet{2010A&A...516A..95H}.
Since these do not deviate from the orbital parameters established
by \citet{2009arXiv0907.5205W}, we do not consider it necessary to
delve in to this finding. We also fit for the impact parameter and
$R_{p}/R_{*}$. However, both values as well as the uncertainties
associed with them are extremely skewed. We therefore conclude that
fitting our transit for more than four parameters leads to large inaccuracies
in our model and only report those parameters fitted with the constant
constraints from \citet{2009arXiv0907.5205W} and \citet{2010A&A...516A..95H}.

\begin{table*}[htbp]
 {\scriptsize }\begin{tabular}{|l|c|c|c|c|c|c|}
\hline 
 & \multicolumn{6}{|c|}{{\scriptsize Constant Parameters}}\tabularnewline
\hline 
 & {\scriptsize $R_{*}$ } & {\scriptsize $M_{*}$ } & {\scriptsize $P$ } & {\scriptsize $a$ } & {\scriptsize $e$ } & {\scriptsize $\omega$ }\tabularnewline
\hline 
{\scriptsize Hebrard et al. 2010 Parameters$^{FH}$ } & {\scriptsize $1.007R_{Sun}$ } & {\scriptsize $1.01M_{Sun}$ } & {\scriptsize 111.4367 days } & {\scriptsize 0.455 AU } & {\scriptsize 0.9330 } & {\scriptsize $300.77^{\circ}$ }\tabularnewline
\hline 
{\scriptsize Winn et al. 2009 Parameters$^{FW}$ } & {\scriptsize $0.968R_{Sun}$ } & {\scriptsize $1.05M_{Sun}$ } & {\scriptsize 111.4374 days } & {\scriptsize 0.4614 AU } & {\scriptsize 0.93286 } & {\scriptsize $300.83^{\circ}$ }\tabularnewline
\hline
\end{tabular}\caption{The two groups of parameters that were held constant while fitting.
Fitting runs using the \citet{2010A&A...516A..95H} parameters are
denoted with $^{FH}$while best-fit results calculated with the \citet{2009arXiv0907.5205W}
parameters are designated with $^{FW}.$\label{table:constantparameters}}
\end{table*}

\begin{table*}[htbp]
\begin{tabular}{|l|c|c|c|c|c|}
\hline 
 & \multicolumn{5}{|c|}{{\scriptsize Fitted Parameters}}\tabularnewline
\hline 
 & {\scriptsize $R_{P}$ } & {\scriptsize $R_{P}/R_{*}$ } & {\scriptsize $i$ } & {\scriptsize Mid-Transit Time } & {\scriptsize $\chi_{r}^{2}$ }\tabularnewline
\hline 
{\scriptsize $ThisWork^{FH}$ } & {\scriptsize $0.987_{-0.061}^{+0.013}R_{Jup}$ } & {\scriptsize $0.1007_{-0.0062}^{+0.0013}$ } & {\scriptsize $89.283^{\circ}\pm0.024$ } & {\scriptsize $2455210.6461\pm0.00781HJD$ } & {\scriptsize 1.01389}\tabularnewline
\hline 
{\scriptsize Hebrard et al. 2010 } & {\scriptsize $0.981\pm0.023R_{Jup}$ } & {\scriptsize $0.1001\pm0.006$ } & {\scriptsize $89.269^{\circ}\pm0.018$ } & {\scriptsize $2455210.6420\pm0.0010HJD$ } & \tabularnewline
\hline
\hline 
{\scriptsize $ThisWork^{FW}$ } & {\scriptsize $0.905\pm0.032R_{Jup}$ } & {\scriptsize $0.0961\pm0.0034$ } & {\scriptsize $89.346^{\circ}\pm0.022$ } & {\scriptsize $2455210.6439\pm0.00773$$HJD$} & {\scriptsize 1.10136 }\tabularnewline
\hline 
{\scriptsize Winn et al. 2009 } & {\scriptsize $0.974\pm0.030R_{Jup}$ } & {\scriptsize $0.1033\pm0.0011$ } & {\scriptsize $89.324^{\circ}\pm0.029$ } & {\scriptsize $2455210.6590\pm0.0050HJD$ {*} } & \tabularnewline
\hline
\end{tabular}

{\scriptsize {*}Predicted result}{\scriptsize \par}

\caption{This Work$^{FH}$ represents best fit values using Hebrard et al.
2010 constraints. This Work$^{FW}$ shows best fit values using Winn
et al. 2009 parameters. The variables determined by the two other
groups are shown for comparison. $\chi{}_{r}{}^{2}$ is shown for
a statistical comparison of the two fits.\label{table:fitparameters}}
\end{table*}

\section{Conclusions}

Employing the optical telescope MOST we observe the January 2010 transit
of HD80606b. Using the fitting routine detailed in \citet{oblateness.2003}
and modifying it for the high eccentricity of this planet's orbit,
we address two separate goals. We first investigate a mid-transit
timing discrepancy discovered by \citet{2010A&A...516A..95H} and
find that our mid-transit time of 2455210.6449 $\pm$0.0034 HJD is
within 1$\sigma$ of \citet{2010A&A...516A..95H} results and 20 minutes
earlier (3$\sigma$ difference) than the predicted value determined
by \citet{2009arXiv0907.5205W}. 

As \citet{2003ApJ...589..605W} have theoretically ruled out the possibility
for a separate planet causing these variations and \citet{ExtrasolarMoons}
discarding the hypothesis for the existence of an exomoon, we think
the transit timing variation is caused by systematic errors depending
on the contrasting methods of ground-based and space-based observations.

We also present an independent set of parameters for the planetary
radius, orbital inclination, and mid-transit time of HD80606b derived
from the MOST photometry. We use the orbital and stellar parameters
determined by \citet{2009arXiv0907.5205W} and \citet{2010A&A...516A..95H}
and hold them constant during the fitting process. We find that our
two sets of results are statistically similar to each other and there
are no unusual discrepancies occurring even with different constraints.
With \citet{2010A&A...516A..95H} parameters, we further determine
that all of our results are within 1$\sigma$ of their results. However,
we also discover that our fitted planetary radius is 2.2$\sigma$
smaller than \citet{2009arXiv0907.5205W} result. 

When fitting for our own independent mid-transit time, we find that
our mid-transit times differ between each other by 3 minutes but remain
less than 1$\sigma$ apart. Furthermore they are 18 minutes earlier
than predicted by \citet{2009arXiv0907.5205W}. We also conclude that
the mid-transit timing issue is not due to the differing in the orbital
or stellar parameters of \citet{2009arXiv0907.5205W} compared to
\citet{2010A&A...516A..95H}. 

As we are currently relying on the observations of only two transits,
we lack the data necessary to draw any sound conclusions regarding
the discrepancies in our results. Future observations, both with ground-based
and space-based telescopes, are a necessity in order to better address
the transit timing issue. These observations will also further constrain
the orbital, stellar, and planetary parameters of HD80606b and aid
in bettering our understanding of the formation and evolution of this
unique system.

\acknowledgements{}

We acknowledge funding from the NASA MOST Guest Observer Program,
grant number NNX10AI84G.

\end{document}